\documentstyle[psfig]{aa}

\input epsf         
\begin{document}

\title{Are quasars
 accreting at super-Eddington rates?}

\author{Suzy Collin\inst{1}, Catherine Boisson\inst{1}, Martine 
Mouchet\inst{1,2},
 Anne-Marie Dumont\inst{1},\\
S\'everine Coup\'e\inst{1},
Delphine Porquet\inst{3},
 Evlabia Rokaki\inst{4}}

\offprints{Suzy Collin (suzy.collin@obspm.fr)}

\institute{$^1$LUTH, Observatoire de Paris, Section de
Meudon, F-92195 Meudon Cedex\\
$^2$Universit\'e Paris 7 Denis Diderot, F-75251 Paris
Cedex 05\\
$^3$SAp, Centre d'\'etudes de Saclay, Orme des Merisiers, F-91191 
Gif-Sur-Yvette\\
$^4$Section of Astrophysics, Astronomy \& Mechanics, University of
Athens, 15784 Zografos, Athens, Greece}

\date{Received : / Accepted : }

\titlerunning{Are quasars accreting at Super-Eddington rates?}
\authorrunning{S. Collin et al.}

\abstract
 {In a previous paper, Collin \& 
Hur\'e (2001), using a sample of Active Galactic Nuclei (AGN)
 where the mass has been determined by 
reverberation studies (the Kaspi et al. 2000 sample),  
 have shown that if the optical luminosity is emitted by 
 a steady accretion disc,
it implies that about half of the objects of the sample are accreting 
close to the Eddington rate or at
super-Eddington rates.
 We discuss here this problem in more details, 
evaluating different uncertainties, and we conclude that this result is 
unavoidable,
unless the masses are strongly underestimated by reverberation studies. 
This can occur
if the broad line region is a flat
 thin rotating structure with the same axis as the accretion disc, 
close to the line of sight. However the masses deduced 
from reverberation mapping in AGN follow the same
correlation between the 
black hole mass and the bulge mass as normal galaxies  (Laor 2001),  
suggesting that they are correct within factor of a few. There 
are then three issues to the problem: 1.  accretion
 proceeds at Eddington or super-Eddington rates in these objects
 through slim or thick discs; 2.  
 the optical luminosity is not produced directly by the 
 gravitational release of energy, but by another additional mechanism, so 
 super-Eddington rates are not required;
 3. accretion discs are completely ``non 
 standard". 
Presently neither the predictions of models nor the observed 
spectral distributions 
  are sufficient 
to help choosing between these solutions. In particular, even for the 
super-Eddington model, the observed optical to bolometric luminosity 
ratio would be of the order of the observed one. \\
In the super-Eddington solution, there is
a strong anti-correlation
 between the observed velocity widths of the lines and the
computed Eddington ratios  (i.e. the accretion rate to the Eddington rate ratios), 
the largest ratios 
corresponding to the narrowest lines, actually to
 ``Narrow Line Seyfert 1" nuclei. For
 the considered sample, the Eddington ratio
 decreases with an increasing
 black hole mass, 
 while the opposite is found if the accretion rate is assumed to be 
proportional to the 
 optical-UV luminosity,
 as it is usually done.  If
 these results are extrapolated to all quasars, it implies that the amount of 
mass locked 
in massive black holes should be larger than presently 
thought.\\
If the Eddington ratio is assumed to be smaller than unity, the optical 
luminosity has to be
 produced by an additional non gravitational 
mechanism. It has to be emitted by a dense and thick medium located at 
large distances from the center (10$^3$ to $10^4$ 
gravitational radii). It can be due to reprocessing of the 
X-ray photons from the central source in a geometrically thin
 warped disc, or in dense ``blobs" 
forming a 
geometrically thick system, which  
can be a part of the accretion flow or can constitute the basis of an 
outflow.\\
The third possibility is not explored here, as it requires 
completely new models of accretion discs which are still not 
elaborated.}


\maketitle

\section {Introduction}

Since the discovery of quasars and Active Galactic Nuclei (AGN), our ideas 
about their radiation mechanisms have strongly evolved. Both because the 
first quasars were radio loud objects, and the spectral distribution from 
infrared to X-rays was found close to a power law with a spectral index 
$\alpha$ of unity ($F_{\nu}\propto \nu^{-\alpha}$), the overall spectrum was
 attributed 
to the synchrotron and the synchro-Compton processes even in radio quiet 
objects. The  seventies were therefore devoted to build models accounting 
for non-thermal radiation in the framework of accretion onto a black 
hole. A breakthrough occurred with the discovery in the spectrum of 3C 273
of an optical-UV feature called ``Blue Bump", corresponding to an important 
fraction of the bolometric luminosity. Since this emission does not display 
any line or edge and resembles a 
blackbody, Shields (1978) proposed to identify the emitting medium with an 
accretion disc, whose large optical thickness 
and low temperature could provide the observed spectrum. This idea was 
widely 
accepted, and many works dealt with the determination of the black 
hole masses and the accretion rates required to fit the blue bump observed in the 
spectra of AGNs and quasars (Malkan \& Sargent 1982 and subsequent works). 
However, 
before the end of the eighties, it was still assumed that a non-thermal 
power law continuum was dominating the near infrared and contributing 
to the optical band. This idea has been abandoned after the publication of 
the generic spectrum of radio quiet quasars by Sanders et al. (1989), 
showing the presence of an ``infrared bump", likely due to hot dust 
emission peaking at a few $\mu$m. Since dust sublimates at about 1700 K, 
it cannot dominate the spectrum below 1 $\mu$m. It was thus 
implicitly assumed that the bulk of the emission below 1$\mu$m is provided by 
the disc itself. Meanwhile the soft X-ray excess has been 
discovered (Wilkes \& Elvis 1987), and it was suggested that the blue bump 
(which was then called the ``Big Blue Bump" or BBB to distinguish it from 
the ``small blue bump" made of a combination of FeII lines and Balmer 
continuum from the Broad Line Region - or BLR) extends from the 
optical up to the soft X-ray range (Walter et al. 1994). The extension of 
the BBB in the soft X-ray range was explained by Compton
scattering in the X-ray irradiated ``skin" of the accretion disc (Ross \& 
Fabian 1993, Nayakshin, Kazanas \& Kallman 2000).  Finally the last possible 
``refuge" for 
non-thermal radiation in radio quiet objects was strongly challenged since 
the discovery in the hard X-ray band of a ``hump" and an underlying power 
law (Pounds et al. 1990), this last component having an energy cut off well 
below 500 keV. The origin of this 
emission is likely to be inverse Compton process in a hot corona irradiated by 
UV photons produced by the disc (Haardt \& Maraschi 1991, 1993).  So the 
entire spectrum in radio quiet objects seems now well explained only by 
thermal emission.
 In this picture, the optical continuum is attributed 
 mostly to gravitational release in a ``cold" disc.

In a recent study based on black hole 
masses in AGN determined by reverberation mapping, Collin \& Hur\'e (2001) showed 
that the observed optical-UV luminosity is too large to be due to a 
standard thin accretion disc in many objects. We recall these results below,
 and we perform a more detailed study, trying to estimate
different uncertainties. In Section 
2 we determine the Mass-Luminosity relationship found in the ``thin disc 
model". In
Section 3 several causes of uncertainties in this 
determination are reviewed. In Section 4 we compare 
 the expected 
spectral distribution to the observations. Section 5 deals with the possibility of
 super-Eddington 
accretion rates,  and  
their cosmological consequences. In Section 6 we discuss the 
possibility that the optical-UV continuum might provided by 
non-gravitational release of energy.

\section{The Mass-Luminosity relation in the ``thin disc" model}

A major tool for determining the mass of central black holes (BHs) 
in AGN is
reverberation mapping: the study of
correlated variations of the lines and continuum fluxes gives 
 the size of the BLR, while the measure of their Full Widths at Half 
 Maximum (FWHM) gives the bulk velocity
(see for a review Peterson 2001). It has been used  in several tens of AGN, under the
assumption that the BLR (or at least the part of 
the BLR emitting the lines used in these studies) 
is gravitationally bound to the BH . 
The FWHM is thus related to the mass by the equation;
\begin{equation}
FWHM=q\sqrt{GM\over R(BLR)}
\label{eq-FWHM}
\end{equation}
where $R(BLR)$ is the radius of the BLR, and $q$ is a factor of the order 
of unity, depending on the orbital shape and on the distribution of matter.

Several bias can alter this determination (Fromerth \& Melia 2000, 
Krolik 2001), and we discuss their implications in Section 3. In the rest 
of the paper we assume that the 
masses determined with this method are correct within a factor of a 
few. 

A considerable advantage of knowing the black hole mass is 
that {\it it gives 
 an additional constraint on the computed disc spectrum}, which
 depends now only on the accretion rate (besides the inclination angle, but there are strong hints that 
in Seyfert galaxies and quasars it is relatively small). 
In previous fits of the ``Big Blue Bump", both the mass and the accretion
 rate were free parameters.  

 Collin \& Hur\'e (2001) used the luminosities of 34 AGN 
 covering a range of four orders of 
 magnitudes and their masses
 deduced from reverberation mapping
 by Kaspi et al.
 (2000), to determine the bolometric luminosity of these 
objects, assuming that the luminosity at 5100\AA$\ $ 
is entirely due to a steady thin accretion disc. We discuss here
 in more details the uncertainties of the results and their various implications
 \footnote{we correct also a mistake in the 
 determination of the luminosity (cf. below)}.

Thin discs are generally modelled using the $\alpha$-prescription
 for the turbulent viscosity introduced by Shakura \& Sunayev 
(1973). It can be shown however that, for a geometrically thin disc with a {\it local} 
energy release, the viscosity 
prescription does not influence the optical emission, as long as a local blackbody
 assumption is valid. The radiation temperature $T$ at a 
distance $R$ from a BH of mass $M$ is then equal to the effective temperature:
\begin{equation}
\sigma T^4 = {3GM\dot{M}\over 8\pi R^3} f(R,a),
\label{eq-dissipation}
\end{equation}
where the non-dimensional factor $f(R,a)$ takes into account the boundary conditions, and depends on 
$R$ and on the angular 
momentum per unit mass $a/M$ of the BH (Novikov \& Thorne 1973; Page \&
Thorne 1974). There are two extreme cases: a Schwarzschild BH with 
$a/M=0$, and a maximally rotating Kerr BH with $a/M=0.998$. 
The emitted spectrum at a given wavelength in the visible range, say 
$\lambda$=5100\AA, can then be calculated by integrating  over the disc the value of the Planck
function at this wavelength (actually the emission at a given frequency 
$\nu$ is dominated by 
the emission at a radius $R_{\nu}$ such that $T(R_{\nu})=h\nu/k$, cf. 
Frank, King \& Raine (1992); for the optical band, the radius of this region is 
large, $\sim 10^{2-4}R_G$, with $R_G=GM/c^2$ being the gravitational 
radius). 

The
 flux at a frequency $\nu_o=\nu_e/(1+z)$ observed at Earth is given, 
in the Schwarzschild case, by:
\begin{equation}
\nu_o F_{\nu_o}={1\over Abs(\nu_o)}  {4\pi cos(i)  h \nu_e^4 \over c^2 D^2 (1+z)^2 }
\int_{Rin}^{Rout} {RdR \over exp(h\nu_e/kT)-1},
\label{eq-flux-earth}
\end{equation}
where $\nu_e$ is the frequency at emission, and $D$ is the proper distance of the object 
\begin{equation}
D=c{q_0z+[q_0-1][\sqrt{1+2q_0z}-1]}/q_0^2H_0(1+z) .
\label{eq-D}
\end{equation}
This expression is valid for $q_0>0$ (Weinberg 1972).
$z$ is the redshift, $i$ the 
inclination of the disc axis on the line of sight, $Abs(\nu_o)$ the external
 (galactic) absorption, and $Rin$ (respt. $Rout$) the inner (respt. the outer)
 radius of the accretion disc. 

For a Kerr BH, the 
expression is more complicated, as 
the spectrum is formed closer to the BH and therefore is more influenced by
 gravitational redshift and gravitational lens effects, which are strongly
dependent on the disc inclination
(Cunningham 1975). These effects produce harder spectra for 
higher inclinations. However they do not affect the emission of the remote 
regions, like the optical continuum.

Setting the boundary condition $f(R)=1$ and assuming that the 
disc extends down to the BH and 
is infinite (which holds if $kT(R_{out}) \ll h\nu \ll 
kT(R_{in})$), 
Eqs. \ref{eq-dissipation}
 and \ref{eq-flux-earth} can be reduced to:
\begin{equation}
\nu_o F_{\nu_o} \propto   R_G^2 T_{in}^{8/3} \propto  [M\dot{M}]^{2/3} 
\label{eq-MMdot}
\end{equation}
where $T_{in}$ is the temperature at the inner radius. This expression can 
however be strongly incorrect, for instance in the case of a truncated disc.

There are large deviations to the blackbody spectrum, in 
particular in the EUV band, which is emitted by the inner hottest regions of the 
disc where Compton scattering is important. However the spectrum is 
close to a blackbody in the optical 
 band, because the disc is still 
 optically thick  in the corresponding emitting region 
(this is not the case for larger radii),
 and on the other hand Compton diffusions are negligible
(see for instance the reviews of Koratkar \& Blaes 1999 and Collin 2001). Moreover  Hubeny et al. (2001) 
have studied the influence on the emergent spectrum of 
the metal opacity, and found that it is negligible in the optical band (cf. 
their figure 13). The viscosity parameter $\alpha$ can 
have a larger effect, but nevertheless it is limited to a few tens of
percents in this band. 

It is thus possible to determine
 the accretion rate $\dot{M}$ from
 Eqs. \ref{eq-dissipation} 
and \ref{eq-flux-earth}, knowing $M$ and the observed value of 
$\nu F_{\nu}$ at a given wavelength in the optical range, and assuming a 
given value 
of cos$(i)$.
It is important to realize that {\it this can be done even though the local blackbody approximation 
does not hold at higher or lower frequencies}, as mentioned above.
Since in the thin disc model
 the bolometric luminosity $L_{bol}$ is  
 equal to  $\eta \dot{M} c^2$, where the
efficiency $\eta$ is 0.057 and 0.32 in the Schwarzschild and in the 
maximally rotating 
Kerr cases respectively (Novikov \& Thorne 1973), one can deduce $L_{bol}$
 in these two extreme cases.  Note that
 in the optical band the Kerr and the Schwarzschild spectral 
shapes are identical, though the integrated flux
differs by a factor 5.6 for the same accretion rate (cf. Hubeny et al. 
2000).

We have made this computation for the 34 objects of the Kaspi et al. 
sample for which the BH masses are known, for a Schwarzschild BH. In an aim 
of uniformity, instead of 
the observed fluxes in the rest frame at 5100\AA, we  use 
 the luminosities given by 
Kaspi et al. at 5100\AA, in the rest frame, $L5100$, which corresponds to an 
isotropic emission (so in the following we will call it the
 ``isotropic luminosity"): 
\begin{eqnarray}
\nu_e L_{\nu_e}&=& 4\pi D^2 (1+z)^2 \nu_o F_{\nu_o} Abs(\nu_o) 
\\
\nonumber
&=&  {16\pi^2 cos(i) h\nu_e^4\over c^2}
\int_{Rin}^{Rout} {RdR \over exp(h\nu_e/kT)-1}.
\label{eq-lum-em}
\end{eqnarray}
(Note that Collin and Hur\'e (2001) have erroneously identified $L5100$ 
to the monochromatic luminosity of the disc). 
We use 
also the 
``mean" masses given by Kaspi et 
al. (cf. below the definition of the ``mean" mass). $H_0$ is taken equal to 
50 
km s$^{-1}$ Mpc$^{-1}$ (it means that we multiply $L5100$ of Kaspi 
et al. by 2.25, since 
they use $H_0=75$), and $q_0$ to 0.5. 
The discs are assumed to extend up to 2 10$^4 R_G$, and down to 6$R_G$ for a Schwarzschild 
BH. 

\begin{figure}
\begin{center}
\psfig{figure=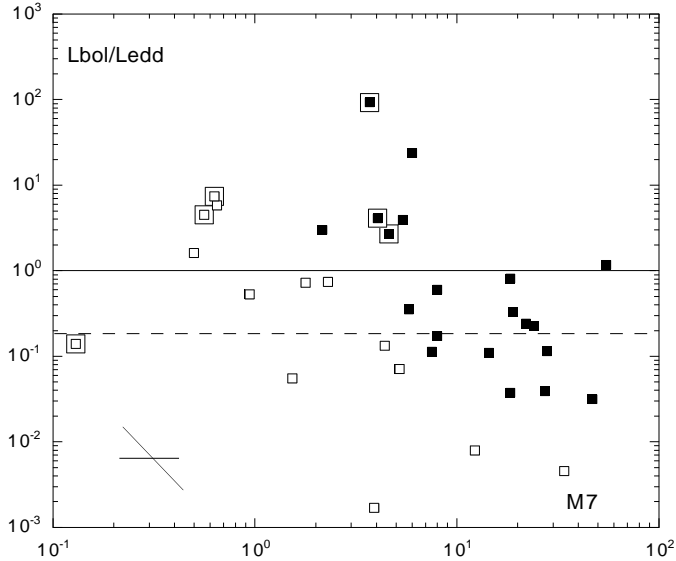,width=9cm}
\caption{Ratio $L_{ bol}/L_{ Edd}$ versus the mass (in units of
10$^7 $M$_\odot$), for the Kaspi et al. sample, assuming that 
 $L5100$ is due to  a thin 
stationary disc 
around a Schwarzschild BH, for cos$(i)$=0.75. Filled squares: $L5100 > 
 2\times 10^{44}$ ergs/s; open squares: $L5100 < 
 2\times 10^{44}$ ergs/s. The big open squares distinguish objects with line
 velocity widths as defined in Section 5 smaller than 1500 km/s.  The solid line 
indicates the Eddington limit for Schwarzschild BHs, and the dashed line 
is the position, with respect to the data points, of the corresponding limit for
 Kerr BHs. The error bars in the left-bottom corner 
give the 
mean error on the mass and the corresponding error on the Eddington ratio.}
\label{fig-Redd-vs-M}
\end{center}
\end{figure}

Fig. \ref{fig-Redd-vs-M} displays the Eddington ratio 
 $L_{bol}/L_{Edd}$ versus the BH mass, 
in the Schwarzschild 
case, for cos$(i)$=0.75. Note that for the same object, 
the Eddington ratio would be larger by a factor 5.6 for a maximally rotating Kerr BH.
We see that even in 
the Schwarzschild case, 11
objects have super-Eddington luminosities, and 7 have luminosities larger 
than 
0.3$L_{Edd}$, the limit of the thin accretion disc. More than half of the sample is therefore not consistent with the model
 of a geometrically 
thin disc, and about one third contradicts the simple idea that super-Eddington 
luminosities cannot exist in the stationary case. In the Kerr 
case, 21 objects (i.e.  more than 
half)
 have super-Eddington luminosities.

\section{What can be wrong with this analysis?}

\subsection{Observational errors}

\subsubsection{Influence of the underlying galactic contribution}

The first idea which comes to mind is that the optical continuum is dominated 
by the 
contribution of the galaxy. It can indeed be the case for low luminosity objects. 
The optical luminosity of 
the disc, and therefore the computed Eddington ratio, would thus be overestimated.

 We have distinguished in 
Fig.  \ref{fig-Redd-vs-M} the objects with a isotropic monochromatic 
luminosity $L5100$ 
 larger than $2\times 10^{44}$ ergs s$^{-1}$.   Except for three  Seyfert 
galaxies which reach this luminosity (Ark 120, Mrk 509 and F9), the other 
luminous objects 
are all PG quasars. Fig.  \ref{fig-Redd-vs-M} shows 
that they form a sequence 
 parallel to local AGN but at higher Eddington ratios.
So the problem is even more dramatic for these objects. However it is unlikely that  
they are dominated 
by the underlying galaxy at 5100\AA. 
Kriss (1988) has estimated the 
stellar fraction in the IR-optical range for a sample of quasars containing 
several
of our PG quasars, and he has found that even for the less luminous 
one this fraction is smaller than 7.5$\%$. Also, a large 
fraction of our PG quasars pertain to the atlas of quasar energy 
distribution of Elvis et al. 
(1994) who give the observed spectrum and the spectrum corrected for the host
 galaxy 
contribution (see their figure 9), and we have checked that for the 
objects of our sample it is always smaller than 10-20$\%$ in the optical 
range. So the problem is not 
solved. 

\begin{figure}
\begin{center}
\psfig{figure=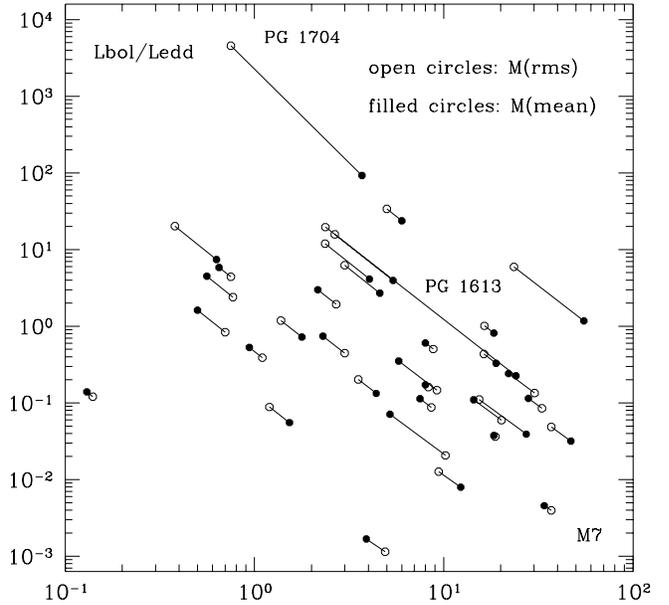,width=9cm}
\caption{Ratio $L_{ bol}/L_{ Edd}$ versus the mass (in units of
10$^7 $M$_\odot$), for the Kaspi et al.  sample, assuming that the
optical
monochromatic luminosity $L5100$ is due to  a thin 
stationary disc 
around a Schwarzschild BH, with cos$(i)=0.75$, using the masses determined by 
the ``mean" (filled circles) and the ``rms" (open circles) methods.}
\label{fig-Redd-vs-M-Mmean-Mrms}
\end{center}
\end{figure}

\subsubsection{Uncertainties on the measured optical luminosities}

According to Eq. \ref{eq-MMdot}, $L_{bol}$ is proportional to 
$L5100^{3/2}$ for a given mass. Thus an uncertainty on the optical flux
induces a larger relative uncertainty on $L_{bol}$.

The optical luminosities in Kaspi et al. (2000) are determined with a good precision (a few 
percents), but they are ``averaged" luminosities which can differ from instantaneous 
ones typically by a factor two, translating in a factor three on 
$L_{bol}$. This uncertainty is smaller than required to suppress the 
super-Eddington rates, and it acts in either directions.

If the objects were preferentially seen face-on (cos$(i)=1$), 
their observed optical luminosity would correspond to a lower value of 
$L_{bol}$
 for a given accretion rate, leading to a decrease of the computed Eddington 
 ratio by a factor 1.5. On the 
 other hand, another 
effect could 
lead to an increase of the Eddington ratio if an object is seen at even 
a relatively small inclination. Due to the large 
amount of matter traversed by inclined light rays, the intensity emitted by the 
disc is larger in the vertical direction if the source function 
decreases like in a stellar atmosphere (it is the
``limb-darkening" effect for the Sun). 
It can increase the influence of the inclination angle, and 
therefore the Eddington ratio for a non-zero inclination.

The luminosity depends on the cosmological constants. The value of the 
luminosity for $H_0$ equal to a value of 
75 km s$^{-1}$ Mpc$^{-1}$ would be 
smaller by a factor two than
for $H_0$ of 50 km s$^{-1}$ Mpc$^{-1}$ chosen here, 
corresponding to a decrease by a factor three of $L_{bol}/L_{ Edd}$. 
 The influence of $q_0$ is actually negligible,
 owing to the relatively small redshifts of the sample. Again this is not a 
 solution to the problem.

\subsubsection{Uncertainties on the derived masses}

The uncertainties on the measured masses are more important, since for
a given value of $L5100$,  $L_{ bol}/L_{ Edd}$ is proportional to $M^{-2}$. 

The most obvious error occurs if the BLR is not 
gravitationally bound to the black hole. The mass would then be 
overestimated and the Eddington ratio underestimated. This effect is opposite 
to what we are looking for. This is indeed possible for the 
region emitting UV lines, such as CIV, which can be outflowing (cf. Done \& Krolik 1996). But 
 Balmer lines which are used here for the reverberation study are 
likely to be gravitationally bound (cf. Peterson \& Wandel 2000, who 
show that the velocity dispersion is Keplerian in NGC 5548).

The shape of the BLR can also be a cause of error. For instance the size of 
the BLR would
 be underestimated, and therefore the mass also underestimated, if 
the BLR would be a cone pointing in the direction of the line of sight.
 Krolik (2001) discusses
 in detail several similar systematic errors in the determination of 
the BH masses by reverberation mapping, due to the 
uncertainties on the shape and on the distribution of the orbits, on
the emissivity distribution, besides those simply due to a bad sampling of 
the observations. 
All these uncertainties lead to systematic errors by factors up to three 
in both directions. They could alter quantitatively but not 
qualitatively  our conclusion, since one order of magnitude error on the Eddington ratio 
would not be sufficient to suppress the problem.

Besides the uncertainty on the factor $q$ in Eq. \ref{eq-FWHM} (Kaspi et 
al. assume $q=\sqrt{3}/2$ which corresponds to an isotropic distribution of 
velocities),  there are several methods for determining the FWHMs. They can
 be measured on the averaged line profile, giving a ``mean'' BH
mass, or they can be
measured on the rms profile, giving a ``rms'' mass. The first method is 
preferred by Kaspi
et al. (2000), and the second one is used by Peterson et al. (1998, 2000). For the 
Kaspi et al. sample the two methods give 
masses differing by factors up to three (ten in one case).
Fig. \ref{fig-Redd-vs-M-Mmean-Mrms} displays the ratios  $L_{ bol}/L_{ 
Edd}$ computed for the masses determined  by 
Kaspi et al. (2000) using the rms and the mean methods. For one object only, 
PG 1613+658, the difference between the two 
measurements is as large as a factor 10, transforming a sub-Eddington
into a super-Eddington 
rate for the rms mass. For another object (PG 1704+608), the Eddington ratio is increased by a factor 30, 
and reaches the enormous value of 5 10$^4$! However the two methods do not give 
results differing 
systematically in the same direction. Note that the Eddington 
ratio is not exactly proportional to $M^{-2}$ for a given $L5100$ as it is
predicted by the approximation of Eq. \ref{eq-MMdot}.

The most important error would occur if the BLR would be a thin flat rotating 
structure constituting the outer regions of the accretion disc, as proposed by 
Dumont \& Collin-Souffrin (1990). In this case the factor $q$ in Eq. \ref{eq-FWHM} is equal to 
sin$(i)^2$ which is of the order of $i^2$ for small inclinations. However this 
angle cannot be very small, as it is limited by the ratio of the scale height 
$H$
to the radius (i.e. to the aspect ratio of the disc). Since the accretion disc is
 gravitationally unstable at the distance of the BLR, it is probably not described
 correctly by a geometrically thin Shakura-Sunyaev model (Collin \& 
Hur\'e 2001). Moreover if the disc is a flat structure, X-ray irradiation by the 
central source would be insufficient to account for the broad line 
intensities. An aspect ratio smaller than unity, say 0.3, is however not excluded.
It could lead to a dramatic effect on the Eddington ratio, which 
is proportional to sin$(i)^4/$cos$(i)^{3/2}$
according to Eqs. \ref{eq-flux-earth} and \ref{eq-MMdot}. The Eddington
 ratio would be 
reduced by a factor $\sim 10^{-2}$, if the object was seen at an 
inclination angle 
smaller than 17$^{\rm o}$.
 Statistically it seems however
implausible that all the ``super-Eddington objects" would be seen at such a 
small inclination angle, which should correspond to less than 5$\%$ of the 
total
number of AGN (and to less than 15$\%$ of
 the number of Seyfert 1 and quasars, 
according to the ``Unifying Scheme"). But it is not impossible that a few 
extreme objects are seen at a very small inclination angle, and are in fact
 close to the Eddington rate.

To conclude this section, it is worth recalling that the BH masses determined 
by reverberation mapping are in agreement with those determined by other 
methods, in particular by the method of the galaxy bulge mass (Magorrian 
et al. 1998).  Laor (1998, 2000, 2001)
  has made an extensive study of the masses of PG quasars and their host 
  galaxies. Extrapolating 
  the correlation between the size of the BLR and the optical luminosity found 
in reverberation studies 
to a sample of PG 
quasars studied with the HST, where both the FWHMs of the H$\beta$ line and the 
luminosities of the host galaxies have been determined carefully, he  
found that the PG quasars  
overlap ``remarkably well" the non-linear 
relation between the 
masses of the BHs and the masses of
the host galaxy bulges. He concludes in particular that ``this overlap 
 indicates that any remaining systematic errors are less than a factor 
2-3" (see also Gebhardt et al. 2000, Ferrarese \& Merritt 
2000, and  Ferrarese et al.  2001). 

\subsection{Theoretical uncertainties}

\subsubsection{Influence of the inner and outer radii} 

If the ``cold" disc is truncated at a large inner radius (for instance by becoming hot),  
 the computed optical luminosity would be smaller for a given 
 accretion rate, especially 
 for low accretion rates and large masses. So the computed Eddington ratio would be 
larger, and idem if the outer 
radius is smaller than 2 10$^4R_G$ (the limit of the disc in our computations).
 If on the contrary the disc extends further away, an 
amount of optical radiation could be added, but only small. Moreover
 2 10$^4R_G$  corresponds to the 
radius where the self-gravity of the disc begins to dominate on the  
vertical 
gravity of the black hole, and the disc becomes gravitationally unstable
(cf. Collin \& Hur\'e 2001). The radius of gravitational instability is 
even smaller  for larger masses. So an outer radius of 2 10$^4R_G$
corresponds most probably to an underestimation of the Eddington ratio. 

\subsubsection{Non-stationary discs}

The previous discussion was based on the assumption that the accretion rate is 
constant along the radius. But this might not be the case.

 First, the disc can be non-steady, owing to the existence 
of instabilities such as the  thermal-viscous instability
(cf. Siemiginowska, Czerny, \& Kostyunin 1996). For these authors, the active 
phase of a galactic nucleus
 corresponds to outbursts induced in the disc by the instability. In this case
 the optical 
luminosity is not necessarily related to the instantaneous accretion rate in the 
same way as in 
steady discs.  However it is difficult to make detailed predictions of the 
spectrum, which depends on the viscosity parameter $\alpha$. 

The accretion rate can also decrease with a
 decreasing radius. It can be the case if a fraction of the 
 accretion flow is converted at small radii into an outflow.
Such flows should be observed through Broad Absorption Lines and/or X-ray 
absorption. Only one object of the super-Eddington ones is a BAL 
quasar (PG 1700+518), a fraction comparable to the average fraction of 
BALs among all quasars. We will see in the next section that no strong X-ray 
absorption seems to be present in the super-Eddington objects. 
So this assumption is unlikely.

\bigskip
Finally if one allows for a fraction of the accretion power to be 
dissipated in a corona and not in the disc itself, the problem of the 
optical luminosity will be even worse, since only a fraction of this 
energy will be returned to the disc in the form of X-ray heating.

\bigskip

Unless the masses determined by reverberation studies are in strong 
failure, it is thus difficult to avoid the conclusion that for a large
 fraction of the 
objects the geometrically thin disc model is not valid.
We are faced to the solutions: 

\begin{itemize}

\item  the accretion rate is really super-Eddington; it is therefore necessary 
to appeal to ``slim" or `` thick" discs, which have a low efficiency $\eta$ 
for converting the gravitational energy into radiation; in this case one has to
 consider the consequences on quasar evolution.

\item  the optical luminosity is due to an additional component, not directly 
provided by the accretion disc, and the accretion rates are all 
sub-Eddington.

\item  the accretion disc is completely different from a ``standard" one.

\end{itemize}

\section{Spectral distribution predicted for accretion discs}

We first examine whether the disc model 
can give a spectral distribution in agreement with the observations, and 
therefore help to discriminate between the solutions.
If both the sub and the 
super-Eddington objects are accounted for by disc emission, we have to 
consider 
not only thin, but also slim 
and thick disc models. 

\subsection{Slim discs or geometrically thick discs}

Geometrically thick discs have been introduced in the framework of the 
$\alpha$-viscosity prescription and
 discussed by Abramowicz et al. (1980). They are characterized
 by an
important advective cooling term, which induces a saturation
 of the luminosity 
for super-Eddington accretion rates. 
In the following we shall therefore use the 
term ``Eddington ratio" for the ratio $\dot{m}=\dot{M}/\dot{M}_{Edd}$ instead of 
$L_{bol}/L_{Edd}$ (where $\dot{M}_{Edd}=L_{Edd}/\eta c^2$). Note that for a Kerr BH, $\dot{m}$ must be multiplied by 
5.6, according to our definition of $\dot{M}_{Edd}$.
Madau (1988)
has shown that the existence of a central funnel leads to a strong 
influence of the viewing 
angle on the spectrum: the overall spectral distribution is softer when seen at
 large inclinations.
 However the 
influence of the inclination on the optical continuum should be quite similar
to the thin disc.  

Slim discs (Abramowicz 
et al. 1988) are a 
vertically averaged version of the thick disc,  valid up to 
 Eddington ratios of a few. The spectra of slim discs have 
been more studied than those of thick discs
 (Szuszkiewicz et al. 1996, 
Wang et al. 2000, 
Mineshige et al. 2000). They differ from those of thin discs mainly in the 
UV and soft X-ray ranges: they are more influenced by Compton 
scattering which results in a higher cutoff energy, and by advection 
which produces a flattening of the spectrum. Slim or thick discs are more parameter 
dependent than thin discs. In 
particular their viscosity parameter $\alpha$ is fundamental in determining their 
structure, which in turn allows to compute the spectrum (a very small 
value of $\alpha$ is generally used, to avoid the disc becoming optically 
thin). They are also more sensitive to the inclination. However the
structure of the outer regions of the disc is not altered with respect to 
a thin disc, so the spectrum and the flux are correctly described 
with the thin approximation in the optical range
 (Mineshige et al. 2000).

\subsection{Comparison of the observed and predicted optical-UV continuum}

\begin{figure*}
\begin{center}
\psfig{figure=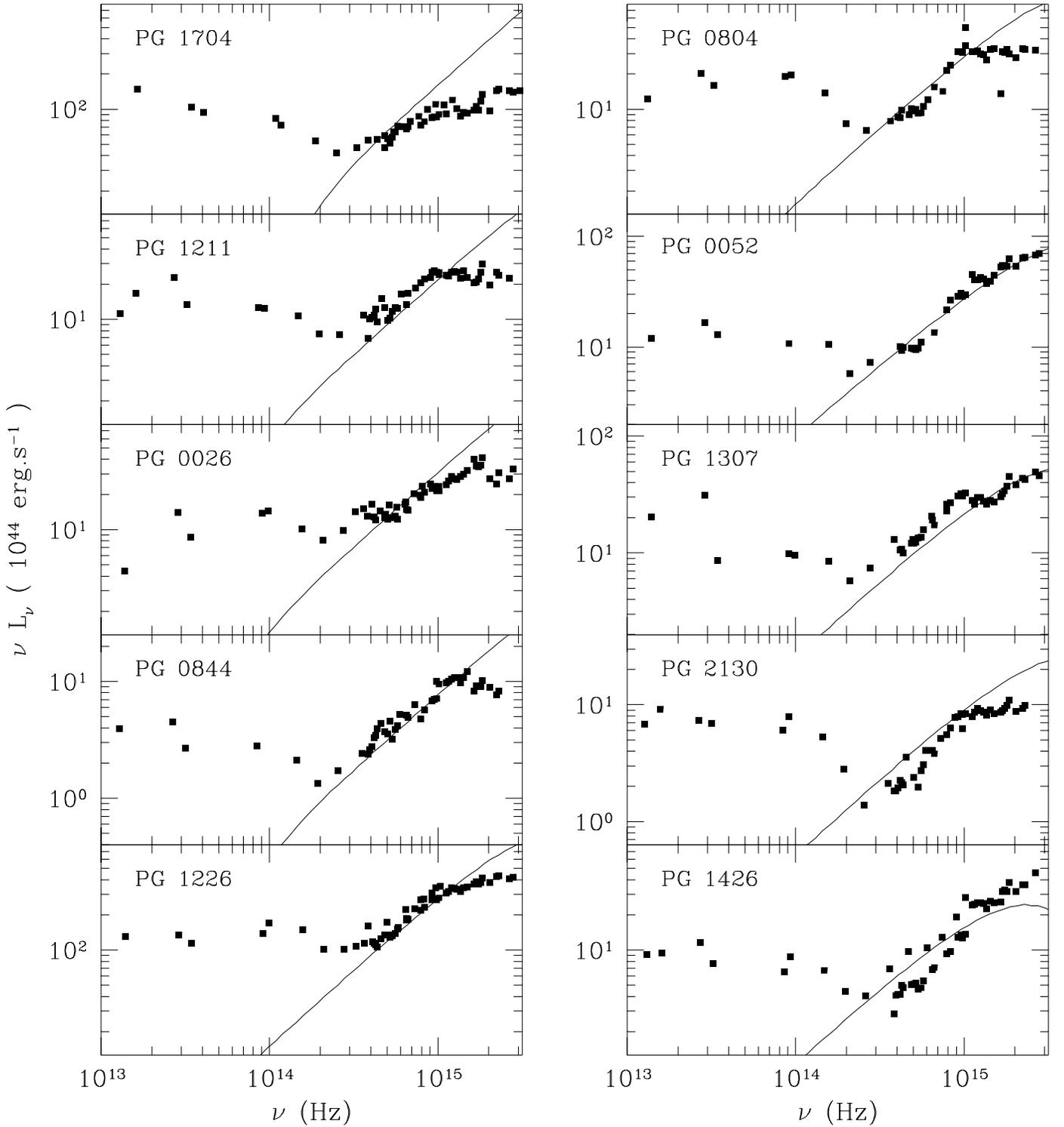}
\caption{Rest-frame, deredened, host galaxy starlight subtracted,
IR-Opt-UV
energy distributions of the
10 PG quasars of our sample from the Quasar Energy Distributions Atlas
(Elvis et al. 1994). The 5 objects on the left have a super-Eddington rate
for a  Schwarzschild BH.
Full line: computed accretion disc spectra derived from the masses and
optical 5100\AA$\ $fluxes as in  Kaspi et al. Note the departure from the
disc
spectra in the IR, which is generally attributed to the dust contribution.}
\label{fig-spectres}
\end{center}
\end{figure*}

It is thus possible to determine the spectrum
 of the disc, assuming 
that it extends far enough to emit the optical continuum. In this case the disc 
 continuum has a spectral distribution 
$F_{\nu}\propto \nu^{+1/3}$ (even in the case of a thick disc). 

We have 11 objects in common with the Quasar Energy Distributions Atlas of
 Elvis et al. (1994) who have gathered the 
near IR-optical and UV observations from different sources. They have corrected their spectrum 
for reddening and for the host galaxy contribution. Figs. 
\ref{fig-spectres} 
 compare the disc emission and the observed IR to UV 
continuum of 10 of these 11 PG quasars. The continuum at 5100\AA$\ $ differs 
slightly from the values given by Kaspi et al (2000) used here to deduce 
$\dot{M}$, 
because the host galaxy contribution was not removed in Kaspi et al. (but it 
amounts to less 
than 20\%), and the observations have been made at different epochs (we 
have actually suppressed one object of the Elvis et al. sample, PG 1613+658, 
where the two values of $L5100$ differ by almost a factor two). The objects 
are ordered by decreasing value of the Eddington ratio from top to bottom 
and from left to right. 

 One sees first that the disc 
continuum 
decreases very rapidly in the IR band, which is generally attributed to the
 dust contribution. We have checked that all the 
objects can be fitted correctly between 3$\mu$m and 3000\AA$\ $ (i.e. 
between 10$^{14}$ and 10$^{15}$ Hz) by a mixture 
of dust and disc emission.
A similar result would be obtained with an additional
 power law 
component $F_{\nu} \propto \nu^{-1\ {\rm to} \ 1.5}$ instead of dust 
emission. 

Second we notice that the computed spectrum is too hard in the UV range, as already 
stressed by several authors (cf. for instance Koratkar \& Blaes (1999) review).
 It seems that the 
``sub-Eddington objects" are better fitted by the accretion disc model.  A way to 
reconcile the disc model with the spectrum of the super-Eddington objects is to
 assume that the ``cold" disc 
is truncated for $R \le R_{in}$, with $R_{in}$ being much larger than the 
radius of the last 
stable orbit, giving rise for instance to a hot corona.

\subsection{Predicted X-ray continuum and X-ray luminosity}

 At least in the case of slim discs, one expects that the emergent spectrum
 should extend
into the soft X-ray range (Szuszkiewicz et al. 1996, Wang et al. 1999),
corresponding to a strong soft X-ray excess.

In order to study this effect in more detail, we have 
gathered the results concerning the 2-10 keV  and
0.5-4.5 keV fluxes and luminosities from different sources, restricting
 our study to the most luminous objects, i.e. the PG 
quasars. The data
constitute an inhomogeneous set, observed at different epochs and 
reduced with different assumptions on the spectral shape and on the X-ray
absorption, but nevertheless they give qualitative information.

Table 1 summarizes the data. Column 1 gives the name of the object, column 
2 the redshift, column 3 the mass in 10$^7$ M$_{\odot}$, column 
4 the isotropic luminosity at 5100\AA$\ $(rest frame), 
 column 5 the line velocity computed as 
$V=\sqrt{GM/R(BLR)}$, column 6 and column 7, respectively the 
 2-10 keV and the 0.5-4.5 keV isotropic luminosities, 
column 8 the Eddington ratio $\dot{m}$ in the case of a Schwarzschild BH and an 
inclination cos$(i)=0.75$. Columns 3 to 5 are derived from Kaspi et al. 2000.

\begin{table*}
\begin{center}
\psfig{figure=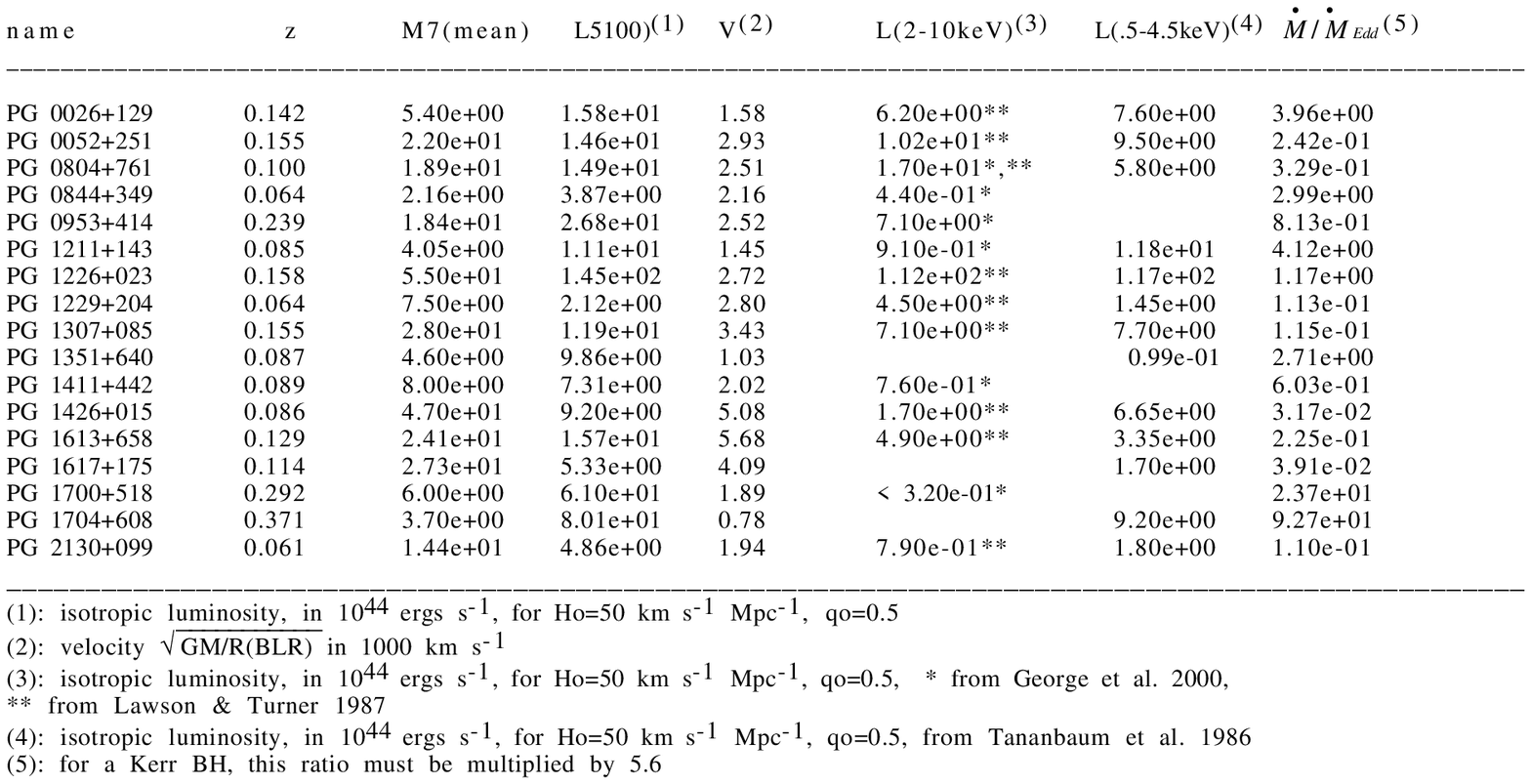}
\caption{Observed properties of the PG quasars from 
the Kaspi et al. sample, and Eddington ratios.}
\label{Table-PG}
\end{center}
\end{table*}

\begin{figure}
\begin{center}
\psfig{figure=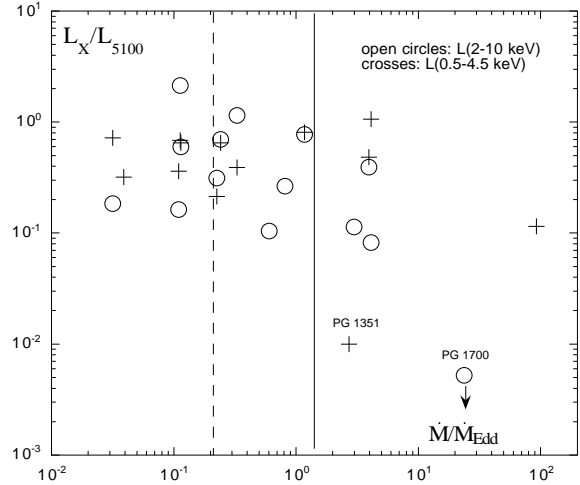,width=9cm}
\caption{L(0.5-4.5keV)/L5100 and L(2-10keV)/L5100 versus the Eddington 
ratio for the PG quasars of the Kaspi et al. sample.   The solid line 
indicates the Eddington limit for Schwarzschild BHs, and the dashed line 
is the position, with respect to the data points, of the corresponding limit for
 Kerr BHs.}
\label{fig-LXsL5100-vs-LbolsLedd}
\end{center}
\end{figure}

Fig. \ref{fig-LXsL5100-vs-LbolsLedd} displays the ratios 
$L(0.5-4.5$keV)/$L5100$ and $L(2-10$keV$)/L5100$ versus the Eddington ratio. 
Two objects, PG 
1700+518 and PG 1351+640, have very low X-ray to optical ratios.  PG 
1700+518
 is a BAL quasar, and these objects are well known to be X-ray weak.  
PG 1351+640 is one of the super-Eddington objects, and it has a strong 
soft X-ray excess (Rush \& Malkan 1996). It also exhibits UV band 
absorption features, however not so broad as a bona fide BAL QSO (Brandt et 
al. 2000). The other super-Eddington 
objects do not seem to
display any particular
 behaviour in the X-ray range: 
they appear to have both normal $L_X/L_{opt}$ and 
$L_{(soft-X)}/L_{(hard-X)}$ ratios, 
showing neither a strong soft X-ray excess, nor a strong X-ray absorption. 
From the very small number of objects for which both soft and 
hard X-ray data are available, it is therefore not possible to draw any 
conclusion. 

\subsection{Optical to bolometric luminosity ratio}

\begin{figure}
\begin{center}
\psfig{figure=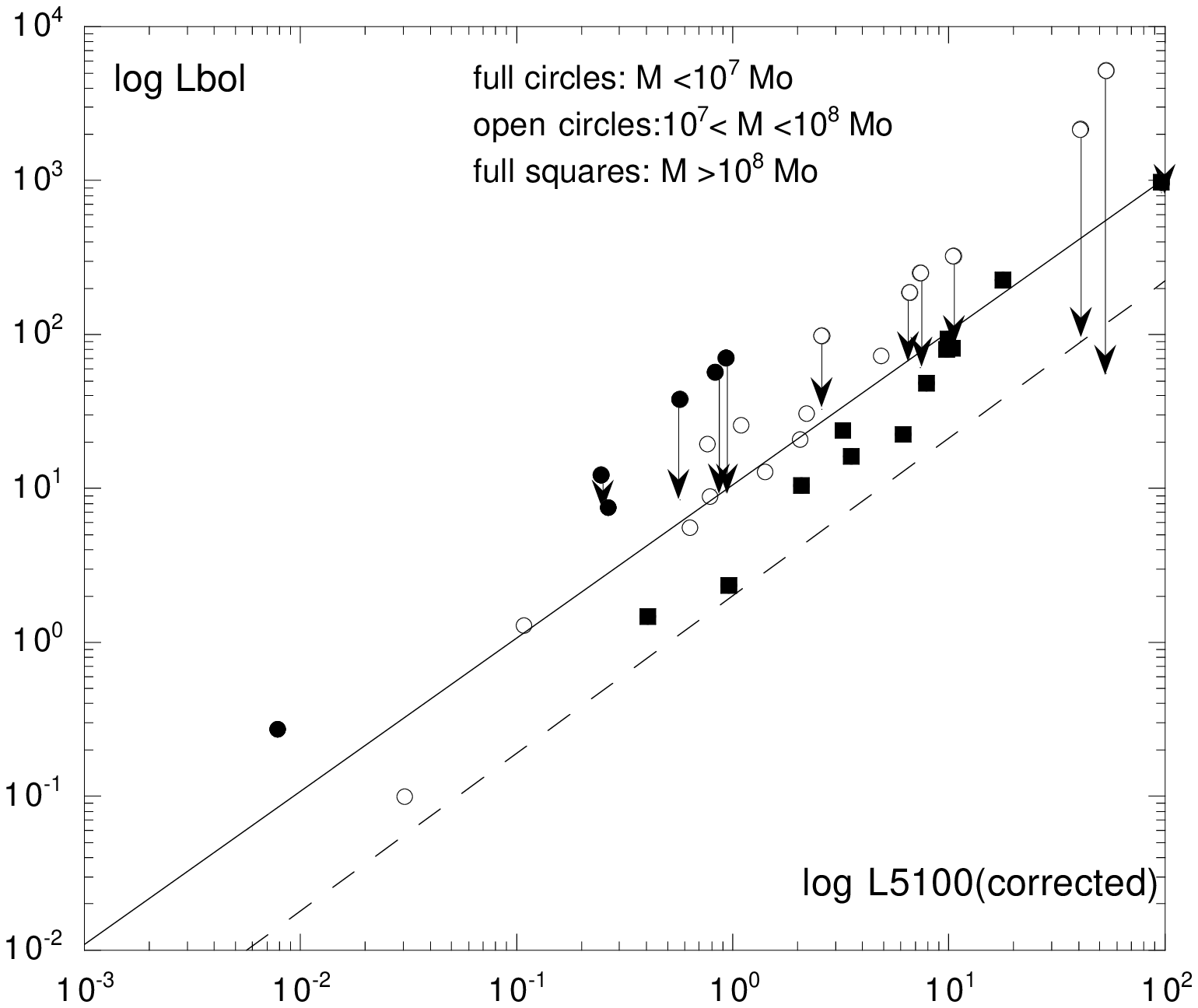,width=8.5cm,angle=0}
\caption{$L_{ bol}$ versus $L5100$(corrected), i.e. divided by 2cos($i)$
 as explained in the text,
in 10$^{44}$ ergs s$^{-1}$
for the Kaspi et al. sample, assuming that 
$L5100$ is due to  a thin 
stationary disc with cos$(i)$=0.75,
around a Schwarzschild BH. We have distinguished three ranges of 
BH masses. Solid line: relation $L_{bol}\sim 10 \times L5100$, for a 
Schwarzschild BH; dashed line: corresponding relation 
 for a Kerr BH. The arrows 
indicate the change in $L_{bol}$, if $L_{bol}/L_{Edd}$ is set equal to 
unity in the super-Eddington objects. }
\label{fig-Lbol-vs-L5100}
\end{center}
\end{figure}

A question one can ask is whether the observed ratio 
$L_{bol}/L_{opt}$ agrees roughly with the computed ratio. The observed 
 generic spectrum of AGN (Laor et al. 1997) corresponds to
 $L_{bol}/L_{opt}$ of the order of 10. To compare the isotropic 
 luminosity $L5100$ to the 
 computed bolometric luminosity, it is necessary to divide $L5100$ by a
factor 2cos($i)$, if the disc radiates like a local 
blackbody at all frequencies, since:
\begin{equation}
L_{bol}= {8\pi^2  h\over c^2} \int_0^\infty \nu_e^3 d\nu_e
\int_{Rin}^{Rout} {RdR \over exp(h\nu_e/kT)-1}.
\label{eq-Lbol}
\end{equation}
Note that large
uncertainties plague the determination of the observed spectral distribution 
in the UV and EUV ranges 
 (the inclination in the Kerr case, the influence of Compton scattering...). 

Fig. \ref{fig-Lbol-vs-L5100} displays $L_{bol}$ versus $L5100$(corrected),
for a Schwarzschild BH and  for an inclination cos$(i)=0.75$. A quick look shows
 an apparently good correlation between the 
two variables, corresponding to $L_{bol}\sim 10 \times L5100$, as observed. 
 But when different 
ranges of BH masses are considered,  the 
correlation is split: for a given value of $L5100$, $L_{bol}$ is 
smaller for a larger mass. 
For a given value of $L5100$, Eq. \ref{eq-MMdot} shows indeed that
$\dot{M}$ (i.e. $L_{bol}$), is inversely proportional to $M$. As we see on 
the figure, 
$L_{bol}\sim 50\times L5100$ for a few objects having $M\le 10^7$M$_{\odot}$, and
  $L_{bol} \sim (2-3)\times L5100$ for $M\ge 10^8$M$_{\odot}$. In the case of a Kerr 
BH,  $L_{bol} \gg 10\times L5100$, in strong disagreement with the observed 
spectral distribution. Note that the actual ``observed" bolometric luminosity 
should be even larger than given here, owing to the increased amount of EUV 
photons due to the inclination in the Kerr case, and to Compton 
scattering, in either the Kerr and the Schwarzschild cases.

However Fig.\ref{fig-Lbol-vs-L5100} does not take into 
account the fact that the $L_{bol}$ cannot be larger than $L_{Edd}$.  
According to the thick disc model, when the accretion 
rate is larger than the Eddington rate, the 
radiative
efficiency $\eta$ is low. Actually the product $\eta \times \dot{m}$ 
should be about 
constant in order to satisfy the relation $L_{bol}/L_{Edd} \sim 1$. 
In Fig. \ref{fig-Lbol-vs-L5100} the 
super-Eddington objects
are shown according to this limit. One sees that the relationship 
between $L_{opt}$ and 
$L_{bol}$ is now in better agreement with the observed one $L_{bol}\sim 10 L_{opt}$.

\bigskip

In conclusion of this section, we see that the ``super-Eddington solution" 
is not dismissed by the observations. 
The optical-UV spectrum can be accounted for 
by an accretion disc, provided that the flux in the UV range is decreased with 
respect to the thin disc solution in 
super-Eddington objects. Also in the X-ray range there are neither enough 
observations and nor sufficiently 
high spectral resolution data
to  disentangle the possible combined effects of absorption and 
soft X-ray excess. 

\section{First possibility: super-Eddington accretion rates}

\subsection{Correlations}

 The spectral and variability properties exhibited
 by  Narrow Line Seyfert 1
Galaxies (NLS1s) have led to think 
that they could be radiating close to their Eddington luminosity (cf.
 Boller et al. 2000). Therefore slim discs have been suggested to 
account for their large ``soft 
X-ray excess" (Mineshige et al. 
2000). 
Like in Collin \& Hur\'e (2001),  in Fig. \ref{fig-Redd-vs-M} we have distinguished
the NLS1s (defined here by
$V = \sqrt{GM/R_{ BLR}} \le 1500$ km/s, with $M$(mean) and $R_{ BLR}$ 
given by Kaspi et al., instead of 
 the usual definition: FWHM$\le 2000$ km/s).
This figure
confirms that all except one NLS1 are super-Eddington objects (the 
exception being NGC 4051, which is peculiar in having a very low mass).

\begin{figure}
\begin{center}
\psfig{figure=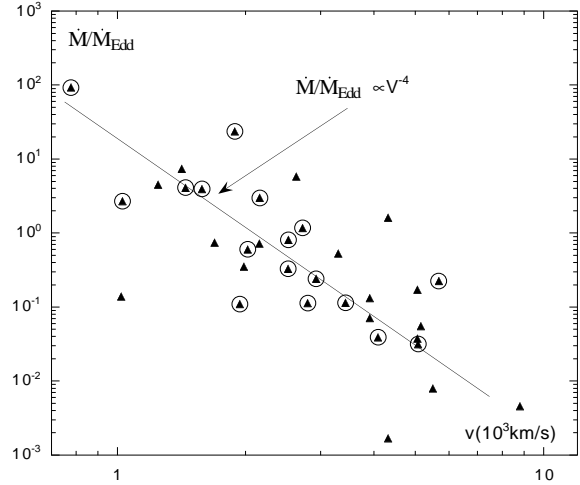,width=9cm}
\caption{Ratio $\dot{M}/\dot{M}_{ Edd}$ versus $V = 
\sqrt{GM/R_{BLR}}$, for the Kaspi et al. sample, for a disc 
around a Schwarzschild BH with cos$(i)=0.75$. Big open circles correspond to 
the PG objects.}
\label{fig-Redd-vs-V}
\end{center}
\end{figure}

However NLS1s do not seem peculiar with respect to the other objects, as is 
shown in Fig. \ref{fig-Redd-vs-V}. When all objects of the Kaspi et 
al. sample are included, one finds that  $\dot{m}$
is roughly proportional to $V^{-4}$. It translates into a relation 
$R_{BLR} \propto \sqrt{M\dot{M}}$. The dependence of the size of the BLR on the
BH mass and on the 
accretion rate could be explained by a combined effect of the radius of gravitational 
instability of the disc and of the ionization parameter, related to $M\dot{M}$, 
as suggested by Collin \& Hur\'e (2001).  Note that the most luminous objects 
(the PG quasars) are not different from the others in this context.

\begin{figure}
\begin{center}
\psfig{figure=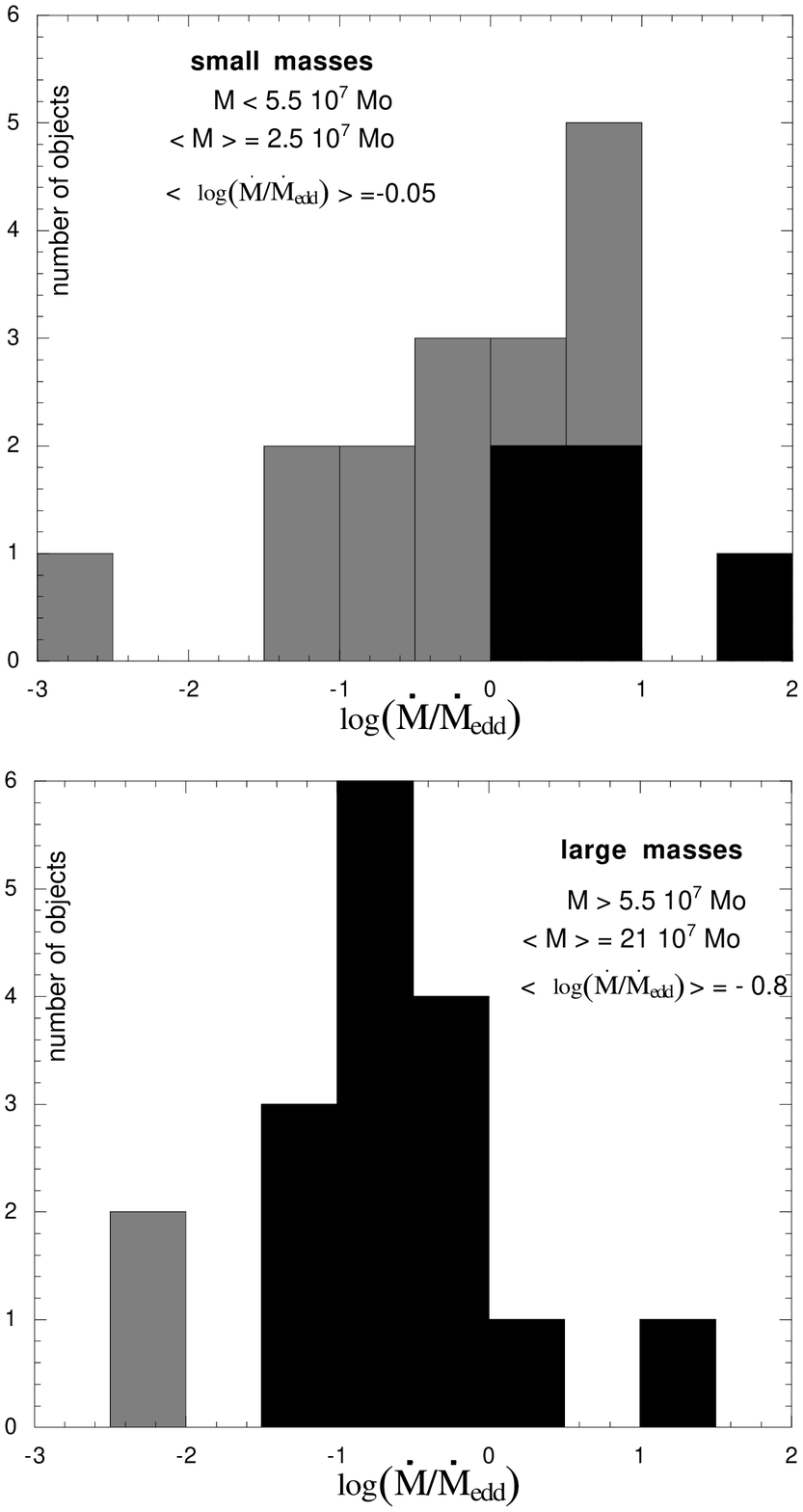,width=9cm}
\caption{Histogram of the ratios $\dot{m}=\dot{M}/\dot{M}_{ Edd}$, for the Kaspi 
et al. sample, for a disc 
around a Schwarzschild BH with cos$(i)=0.7$. The objects with $L5100 \ge 
 2\times 10^{44}$ ergs/s are shown in black.}
\label{fig-hist-Redd}
\end{center}
\end{figure}

A loose anticorrelation between 
 $\dot{m}$ and $M$ appears in Fig. \ref{fig-Redd-vs-M}. Though 
it can be partly accounted for by the 
uncertainties on the masses which act in the same direction, it is 
clear that smaller masses
 correspond to larger 
Eddington ratios. The effect is
 quantified in Fig. \ref{fig-hist-Redd}, where the sample has been split 
 into 
 two equal parts. It shows that objects with an average mass of 
2.5 $10^7$M$_{\odot}$ have an average log($\dot{m})$ equal to=-0.05, while
 objects with an average mass of 2.1 $10^8$M$_{\odot}$ have an average
log($\dot{m})$ equal to -0.8. We have distinguished the objects with $L5100 \ge 
 2\times 10^{44}$ ergs/s. Though their number is small, we see that they display 
the same effect 
 (actually even slightly amplified, but as the sample is too small we 
 cannot be sure that
 it is real), 
meaning that the underlying contribution of the galaxy does not 
introduce any important bias. 

The Kaspi et al. sample is actually made of two parts, on one side 
low luminosity local objects, which have been selected on various criteria, 
including a strong variability, on the other side a sample of 28 PG quasars 
selected on the basis of their positions, redshifts and magnitudes, from 
which only 17 objects had enough observations to be used  
for the reverberation study (Maoz et 
al. 1994, Kaspi et al. 2000). Since we do not know how these 17 objects 
have been selected, we cannot exclude a bias, for instance favoring more 
strongly variable objects or more luminous objects.

Several authors have deduced from their studies a relation 
between  $M$ and 
the optical or UV luminosity. For instance  Wandel, Peterson \& Malkan (1999)
 found 
$M\propto L_{opt}^{0.77}$ and
 Kaspi et al. (2000) found 
$M\propto L_{opt}^{0.5}$. Thus the assumption that  $L_{opt}/L_{ bol}$ is constant
which was always made before, leads to the result that $\dot{M}$
increases with $M$. On the contrary, we find that the Eddington
 ratios are on average one order of magnitude larger for masses one order of 
 magnitude smaller (cf. Fig.  
\ref{fig-hist-Redd}), i.e. {\it that $\dot{M}$ is about inversely 
proportional to $M$}.  As already mentioned, this is due to the fact that
 the ratio $L5100/\dot{M}$ increases 
 with the mass, since larger masses correspond to an emission peaking
 at smaller frequencies for the same accretion rate.

\subsection{Evolutionary consequences}

The fact that the Eddington ratio can be larger than unity
 has important evolutionary implications. A BH accreting at the 
 rate $\dot{m}=\dot{M}_{ bol}/\dot{M}_{ Edd}$ doubles its mass in a time
of $4\ 10^8\eta/\dot{m}$ years. 
If some quasars are accreting with a small efficiency  at a super-Eddington rate,
their growing time is 
reduced by
$\dot{m}^2$, and their ``active 
phase" is therefore very short compared to the Hubble time. In particular it 
is much smaller than the time for the merging or for the interaction of two galaxies. 

Another important cosmological implication concerns the mass 
 in ``dead black holes". The mass 
density 
locked in black holes, required to account for the visible light of quasars, is 
indeed  (cf. Soltan 1982,  Chokshi \& Turner 1992):

\begin{equation}
\rho(QSO)\sim 2\ 10^5 {L_{bol}\over 10 L_{opt}}{0.1\over\eta}\ \  {\rm M}_{\odot}\ 
{\rm Mpc}^{-3}.
\label{eq-mass-density-quasars}
\end{equation}

We have seen that a proportion of quasars, which can amount to half in the 
considered sample,
has a super-Eddington accretion rate, and therefore an efficiency $\eta$ smaller
 than 0.1. It is of course impossible to determine the correction factor 
 it would induce on Eq. \ref{eq-mass-density-quasars} since we do not have 
 a real statistic for the super-Eddington objects.  Moreover other 
 objects (those accreting at very low rate) can have also a very small 
 efficiency (the ``ADAFs", cf. Narayan \& Yi 1995), so the distribution of efficiencies during the 
 growing and active phase is not known. Nevertheless one can expect that 
 $\rho(QSO)$ is larger than $2\ 10^5$ M$_{\odot}$  
 Mpc$^{-3}$  by a few units.

On the other hand the mass in the nuclei of normal local galaxies, which 
should be the remnants of the quasar population, is estimated from the 
Magorrian et al. (1998) linear relation between the BH and the host galaxy bulge 
mass ($M(BH) \sim 0.005 M(bulge)$) by
Haehnelt, Natarajan \& Rees (1998): 
\begin{equation}
\rho(BH)\sim 3\ 10^6 {M_{BH}/M_{bulge}\over 0.006}{\Omega_{bulge}\over 
0.002}\ \  {\rm M}_{\odot}\ 
{\rm Mpc}^{-3}.
\label{eq-mass-density-BH}
\end{equation}
where $\Omega_{bulge}$ is the mass density in bulges.

This estimate has to be revised according to Laor's (1998, 2001) finding 
that the relation between $M_{BH}$ and $M_{bulge}$ is not linear, and  
the ratio $M_{BH}/M_{bulge}$ is about 0.0005 for low luminosity bulges and 
0.005 in bright ellipticals. $\rho(BH)$ is thus probably smaller by one 
order of magnitude and therefore close to, or even smaller than, the 
value of $\rho(QSO)$. This is an uncomfortable result, owing to the fact that 
some late accretion could also take place in local BHs. 

\begin{figure}
\begin{center}
\psfig{figure=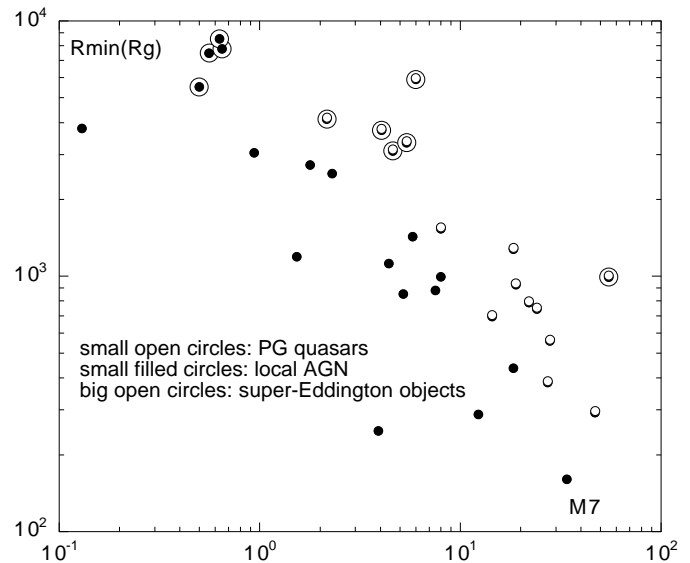,width=9cm}
\caption{Minimum radius of the region emitting the optical luminosity, in 
$R_g$, for the Kaspi et al. sample.}
\label{fig-Rmin-vs-M}
\end{center}
\end{figure}

\section{Second possibility: the Blue Bump is not emitted by 
gravitational release of energy}

\subsection{Physical conditions required for the emitting medium}

The observed optical spectrum of AGN is not featureless, as it contains 
intense emission lines and a Balmer discontinuity. But they are attributed 
to the BLR. The BLR has been modelled 
successfully since two decades by a relatively dilute medium 
(with a density $10^9 \le n \le 10^{12}$ cm$^{-3}$). Its optical thickness 
in the visible range is smaller than unity, and it emits a negligible 
continuum at wavelengths larger than the Balmer discontinuity. Since the observed 
Balmer discontinuity is attributed to 
the BLR, there is no place left for another important contribution. 

Therefore is raised a general problem, namely the ability of 
an emitting medium to give rise to a featureless continuum in the optical 
range. 

Let us now give up the disc model where the dependence of the flux on 
the radius is imposed, and simply use the fact that the medium emitting the 
part of the Blue Bump 
extending in the optical and in the near IR range 
should have a temperature of the order of 
$5\ 10^3-10^4$ K. The {\it minimum radius} of this region, $R_{min}$,
 corresponds to the
 blackbody flux, as any other emission
 would be less efficient in terms of flux
(except non-thermal emission, but we do not consider this possibility here). 
For $T=10^4$K (the temperature corresponding to the maximum of the flux per unit 
frequency at 5000\AA), it is given, within a factor of order of 
unity which depends on the geometry of the emitting medium, by:
\begin{equation}
r_{min}={R_{min}\over R_G} \sim 3\ 10^3 {\sqrt{L5100_{44}}\over M_7}\sqrt{4\pi 
\over \Omega},
\label{eq-radius}
\end{equation}
where $L5100_{44}$ is expressed in 10$^{44}$ ergs s$^{-1}$ 
and ${\Omega\over 
4\pi}$ is the coverage factor of the emitting medium (i.e. its opening 
angle). 

\begin{figure}
\begin{center}
\psfig{figure=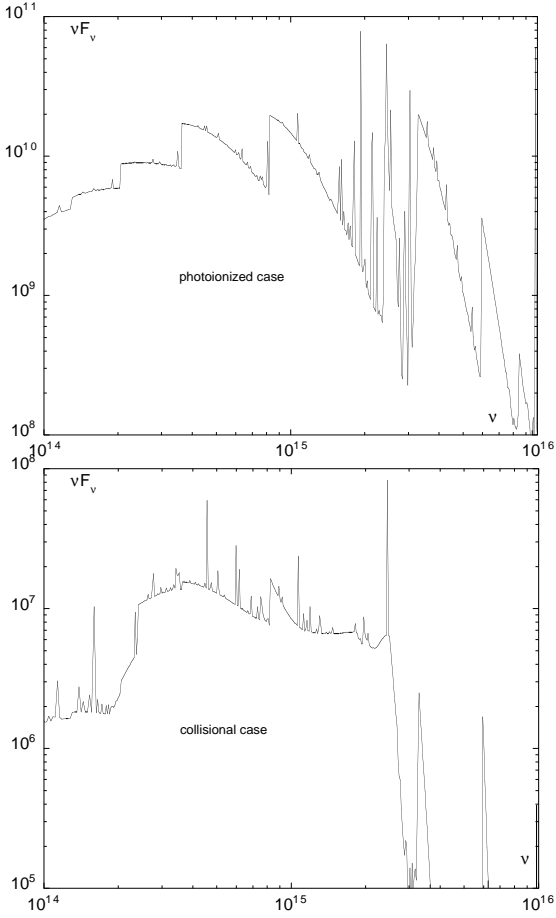,width=9cm}
\caption{Total spectrum (reflected + emitted) for a slab of density
 $10^{12}$ cm$^{-3}$ and column density $10^{24}$ cm$^{-2}$. Top: 
 photoionized case, 
 assuming a photoionizing continuum 
 $F_{\nu}\propto \nu^{-1}$ from 10 eV to 100 MeV, and an ionization 
 parameter $U=10^{-2}$. Bottom:  hybrid 
case of a collisionally 
and radiatively ionized slab with a constant temperature equal to 5000K 
and an ionization parameter $U=10^{-5}$. Computation using CLOUDY 94. 
Spectral resolution: 30.}
\label{fig-Cloudy}
\end{center}
\end{figure}

\begin{figure}
\begin{center}
\psfig{figure=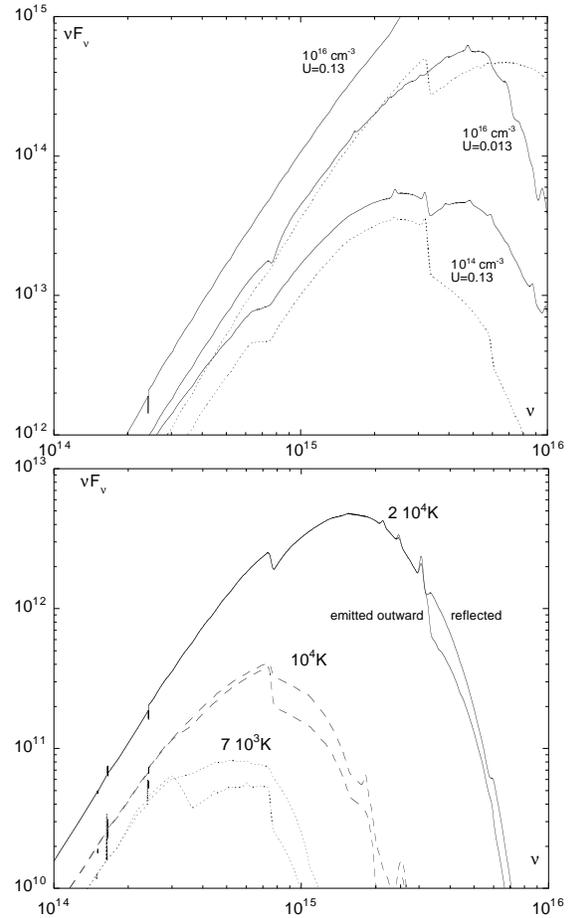,width=9cm}
\caption{Spectra computed using TITAN. Top: 
photoionized slab of density
 $10^{16}$ cm$^{-3}$ and  $10^{14}$ cm$^{-3}$ 
and column density $10^{25}$ cm$^{-2}$. The curves are labelled with the 
density and the ionization parameter.  The shape of the
ionizing continuum is the same as for Fig. \ref{fig-Cloudy}.
 Solid lines: reflected 
spectrum, dashed lines: outward emitted spectrum.
 Bottom:  hybrid cases of a
 collisionally 
and radiatively ionized slab with a constant temperature for a slab of density
 $10^{14}$ cm$^{-3}$ and column density $10^{25}$ cm$^{-2}$. The outward 
emitted and the 
reflected spectra are displayed. The 
temperature 
is indicated on the curves. Spectral resolution: 30.}
\label{fig-Tit}
\end{center}
\end{figure}

\begin{figure}
\begin{center}
\psfig{figure=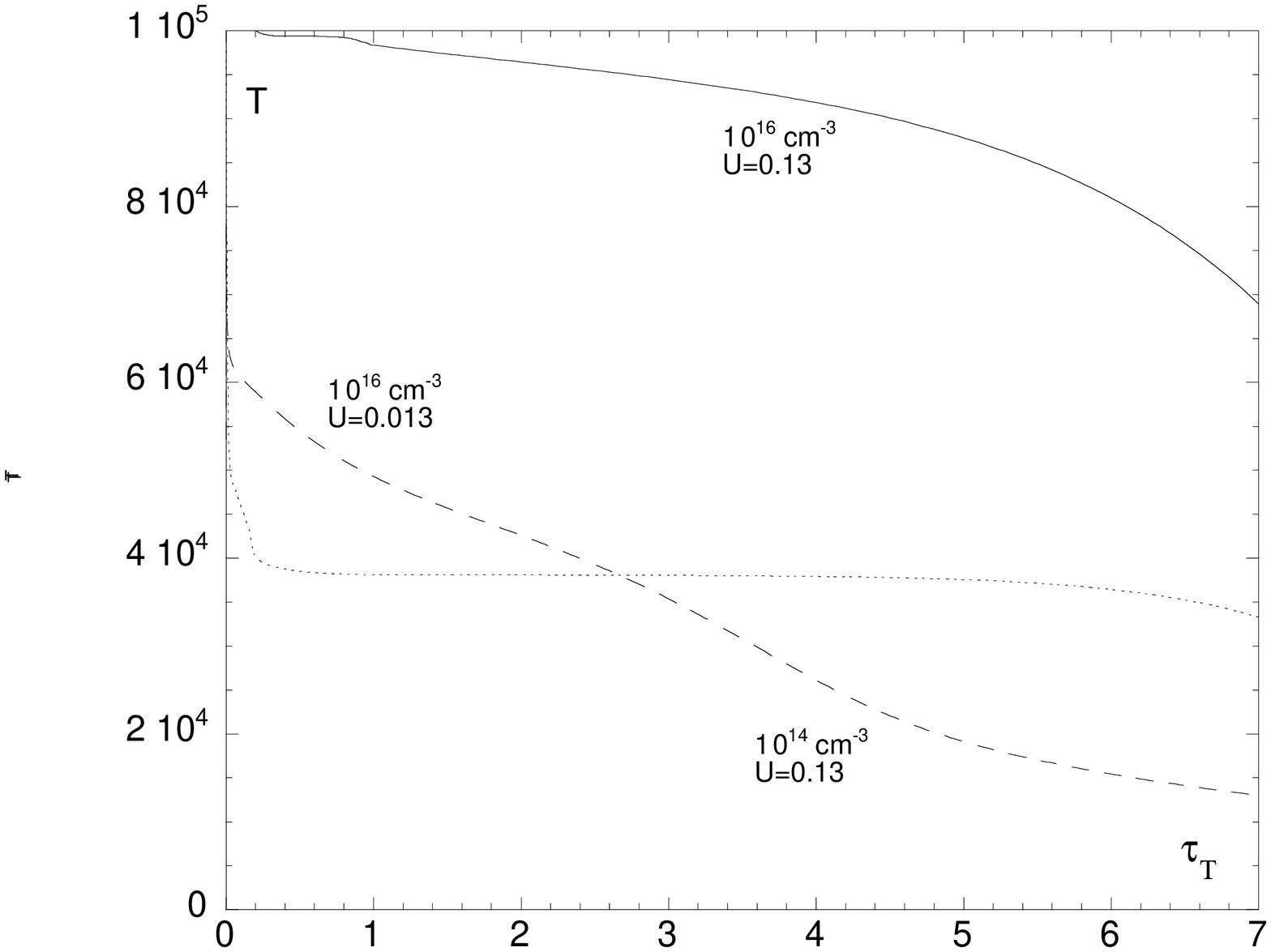,width=9cm}
\caption{Temperature versus the Thomson optical depth for the photionized
models of the 
upper panel of Fig. \ref{fig-Tit}. The curves are labelled with the 
density and the ionisation parameter.}
\label{fig-Tit-T}
\end{center}
\end{figure}

Fig. 
\ref{fig-Rmin-vs-M} displays this minimum radius for all the objects of the sample, 
assuming the maximum plausible coverage factor of 0.5 (corresponding to the largest 
value of $r_{min}$; note that $T=10^4$K corresponds to a spectrum 
peaking at 5000\AA, while a smaller temperature is required for larger 
wavelengths, so $r_{min}$ would be even larger).
 We see that
 the minimum radius for the optical emission is always of the order of or 
larger than a few hundreds $R_g$. In particular it is larger than 10$^3R_G$ for
 all the 
super-Eddington objects. 

Whatever the structure of the emitting medium is, let us assume that 
it is irradiated and it emits by reprocessing of the central X-ray source.

A particularity of photoionized media is that they are generally far from 
Local Thermodynamical Equilibrium (LTE), and consequently their emission 
differs strongly from a Planck spectrum. It contains ionization
 edges and lines which are either in absorption or in emission, depending 
 on the 
 variation of the temperature and of the level populations with depth. To 
 get a featureless optical continuum, a very high density and/or very large
optical thickness (in the optical range) are required.

The medium emitting the optical continuum is at about the same distance as
 the BLR, but it should be much 
closer to LTE than the BLR, and therefore much thicker, 
(i.e. the ionized zone should be optically thick 
 or effectively thick in the optical band). This is 
possible only if the density is much higher than the usual density of the 
BLR. To illustrate this point, the upper panel of Fig. \ref{fig-Cloudy} shows 
the total 
spectrum
(i.e. the sum of the outward and inward spectra)  
 emitted   by a 
photoionized system of clouds with a density $10^{12}$ cm$^{-3}$ and a column
 density $10^{24}$ cm$^{-2}$. The photoionizing continuum is
 $F_{\nu}\propto \nu^{-1}$ from 10 eV to 100 keV, and the ionization
 parameter $U$ defined as:
\begin{equation}
 U={1\over n_Hc} \int_{13.6 {\rm eV}}^{100 {\rm keV}}{F_{\nu}\over h\nu} d\nu
\label{ionizationpar}
\end{equation}
is equal to  $10^{-2}$.
 The spectrum is 
 computed using CLOUDY (cf. Ferland et al. 1998). The incident 
 continuum is cut below 10eV, since we assume that there is no other emission
than that produced by the system of clouds itself  in the optical range (for 
 instance an underlying power law). The emitted spectrum displays 
 very strong Balmer and Paschen discontinuities, incompatible with the 
 observations. The spectrum shown 
on the lower panel of Fig. \ref{fig-Cloudy} corresponds to an ``hybrid" case
 where the medium is both
 collisionally and radiatively ionized, with a constant 
temperature equal to 5000K and an ionization parameter $U=10^{-5}$. This 
spectrum is more
 featureless in the 
optical-near IR range than in the pure photoionization case in terms of 
discontinuities, but nevertheless 
it displays a too strong Balmer 
discontinuity. 

The density and the column density must therefore be larger than $10^{12}$
 cm$^{-3}$ and  $10^{24}$
 cm$^{-2}$ to get 
a featureless continuum. Such 
models cannot be computed using CLOUDY. We have used the code TITAN designed 
for dense Thomson thick media  (cf. Dumont et al.
 2000). Examples of
 such cases are shown in the upper panel of Fig. 
\ref{fig-Tit} which displays the results of computations performed with 
a column density of $10^{25}$ cm$^{-2}$, 
a density of  $10^{16}$ and $10^{14}$ cm$^{-3}$, with an ionization
 parameter $U=0.13$ and $U=0.013$.  The spectrum corresponding to  $10^{16}$ cm$^{-3}$ is closer 
 to the blackbody value, as expected. All spectra are smooth (note however 
 the presence of a few absorption lines) but they peak in the EUV and not in the 
 optical band. This is because the temperature of the emitting shell, which 
 is almost a constant as the medium is close to LTE, is too high for 
 optical emission (cf. Fig \ref{fig-Tit-T}). So one must seek for smaller ionization
 parameters, 
 corresponding to smaller temperatures. 

For practical reasons this is not possible presently with TITAN, which 
does not converge easily in this case; so we 
have computed hybrid cases where the temperature is set as a constant.
 Such cases are shown in the lower panel 
 of Fig. \ref{fig-Tit}, 
 which displays the total spectra in three cases of temperatures, with 
small values of the ionization parameter. We note that the only 
 spectrum peaking in the optical range (corresponding to $T=7\ 10^3$K) 
 displays large Balmer and Paschen discontinuities in its outward 
 spectrum, but not in the reflected spectrum. The respective proportion of 
 the reflected and emitted spectra in the observed spectrum depends on the 
 geometry. It is quite possible that 
 one sees preferentially the reflected spectrum (for instance in a torus 
 geometry, where the observer is located above the torus).  Moreover, 
the cases displayed here constitute only a few examples. We pospone 
to a future paper a more detailed discussion of the problem, which is out 
 of the scope of the present article.

To summarize this discussion, we have seen that, in order to get a spectrum 
peaking in the optical range, and at the same time displaying very small 
edges or lines, it is necessary that the emitting medium, if photoionized, 
has both a large density (10$^{14}$ 
cm$^{-3}$), and a Thomson thickness at least of 
unity. Besides, to get the observed luminosity,
it should be located far 
from the center and it should have a large spatial extension. The question 
is now to identify this medium.

\subsection{Irradiated discs}

It is admitted now that accretion discs are irradiated by the X-ray 
source providing the spectrum observed up to a few hundred keVs in AGN
(see for a review Collin 2001).
This is based on two arguments. First the UV and optical 
emitting bands are located at relatively large distances from each other, 
and the small time lags observed between the  optical and UV 
light curves 
require a causal link propagating at the speed of light. This led to propose 
that the emitting medium is 
irradiated by the X-ray continuum, and emits partly as a result of this
external radiative heating (Collin-Souffrin 1991). Second the presence in 
the X-ray spectrum of 
the iron K fluorescence line and 
of the X-ray hump (Pounds et al. 1990) implies reprocessing by a cold 
medium, likely an optically thick accretion disc.

In the most commonly accepted irradiated disc model, the X-ray source is located
 at a few disc scale heights above 
the inner regions of a geometrically thin disc, either as a ``lamppost", or as a
 patchy corona (the ``flare model"). The irradiated disc absorbs the X-ray 
 photons and 
reprocesses them into 
optical and UV-soft X radiation through atomic processes. In the hard X-ray band, 
the reprocessing is due to Compton diffusion. In the flare model, a 
magnetic flare suddenly 
 releases a large X-ray luminosity, corresponding to a 
larger flux than the underlying viscous one, while it is not the case in 
the lamppost model. 

If the disc is geometrically thin, the regions of the
 disc located at distances from the X-ray source 
larger than the height of the X-ray source (like those emitting the 
optical luminosity) receive a flux decreasing as 
$R^{-3}$, owing to the inclination of the light rays with respect to the 
disc surface (Dumont \& Collin 1990). So the irradiating 
flux decreases like the viscous flux (Eq. \ref{eq-dissipation}). As a consequence the shape of 
the emitted optical continuum ($F_{\nu}\propto \nu^{1/3}$) is not modified with respect
 to the non-irradiated case (contrary to the EUV continuum, since the maximum
 surface temperature reached by the disc 
increases), and the optical 
luminosity cannot be increased by a large factor.

An important fraction of the UV and X-ray luminosity could be 
reprocessed by the outer regions of the disc into optical luminosity 
if the X-ray source is
located at a large height above the disc.
However this assumption contradicts the idea that the 
bulk of the X-rays is produced close to the BH. 

If the scale height of the disc increases more rapidly 
than the radius (the disc is then said to be ``flaring"), the outer 
regions of the disc can be irradiated more efficiently by the central X-ray source.
This explanation has been often proposed  (cf. for 
instance Siemiginowska et al. 1995, Soria \& Puchnarewicz 2002) to account
 for the large amount 
of optical flux, and
for the too ``soft" observed continuum in the optical-UV 
range, compared with that expected from a disc.
However this explanation presents some difficulty for two reasons. 
Either the optical flux is emitted at relatively small distances from the 
center, 10$^2$ to 10$^3\ R_G$, and in this case an $\alpha$-disc in an AGN, which 
is dominated by radiation pressure and Thomson opacity, does not flare
(cf. Shakura \& Sunyaev 1973,  Hur\'e 2000). Or the optical flux is emitted at a
 large distance, 10$^3$ to 10$^4\ R_G$, and in this case the $\alpha$-disc flares, 
 but at the same time its midplane density 
 decreases drastically (in $R^{-33/20}$), and the density is much too small in the 
 irradiated layers to give 
 the required blackbody emission (cf. Dumont \& Collin-Souffrin 1990). 
Note that these reasons do not hold for galactic BHs. First the
 opacity is dominated by atomic processes and not by Thomson scattering, 
 and consequently the structure of the inner regions of the disc is 
 different, and the height over radius ratio increases. Second the density 
 is much larger than for AGN discs.

The flux reprocessed in the optical band can be increased if the X-ray 
photons are back-scattered 
towards the disc, but this requires a Thomson thick medium surrounding the 
disc and the X-ray source. This medium would create a cut-off at low 
energy (a few tens of keV) due to Compton scattering, which does not agree 
with the observed high energy spectrum of AGN. 

 The outer regions of the disc can also be irradiated by the central X-ray 
source if the disc is warped, as suggested by recent observations 
(Kinney et al. 2000). The warping can be due to self-irradiation 
instability (Pringle 1996), to misalignment of the spin of the BH with the disc
 (Baarden \& 
Petterson 1975), or to feeding of the disc through a 
misaligned inflow.
The range of observed spectral index $\alpha$ in the optical band 
($F_{\nu} \propto \nu^{-\alpha}$) is 0 to 0.5. If it is produced as a local 
blackbody by 
reprocessing of an external flux, the integration 
 on the radius leads to an emitted flux $F_{\nu}\propto 
 \nu^{3-2/a}$, where $a$ is defined by the irradiating flux being proportional to 
$R^{-4a}$. 
So the irradiating flux 
should vary as $R^{-8/3}$ for an observed spectral index $\alpha$ equal to 0,
 and as 
$R^{-16/7}$ for $\alpha=$0.5, compared to the non-irradiated 
case where the flux varies as $R^{-3}$. These values are plausible, 
though they imply a large 
warping.

So finally the model of a geometrically thin warped irradiated disc is the
 minimum 
prerequisite to account for the observations. However one should check from 
a detailed modelling, with the observed values of the mass and the 
luminosity, that the required density and column density are reached
 at large distances from the center. 

\subsection{An inhomogeneous emitting medium made of dense clouds}

If the emitting medium is not in the form of a thin disc, its large size
 (typically 
$10^3R_G\sim 10^{16}$cm), combined with a density of at least 
$10^{14}$ cm$^{-3}$, would imply an enormous column density for a uniform 
medium. When located 
on the line of sight, it would completely obscure any emission coming from 
the center, including gamma rays. This is not observed, except perhaps 
in the most 
obscured  Ultra Luminous Infrared galaxies. 

The suggestion that the BBB is emitted by a system of dense clouds 
 reprocessing the X-ray continuum was made by
Celotti et al. (1992), who assumed an emission region made 
of very dense
 (up to $n=10^{20}$ 
cm$^{-3}$) Thomson thin clouds.  Barvainis (1993), Collin-Souffrin et 
al. (1996), proposed an 
emission region made of less dense ($n=10^{12-14}$ cm$^{-3}$) Thomson
 thick clouds. The first model does not account for a featureless 
 continuum in the UV range, where the medium is optically thin 
(cf. Kuncic et al. 1996). The second model can account easily for a 
featureless 
continuum both in the optical and in the UV range. 

An additional advantage of the ``thick cloud" model is that it can 
 also account for the presence of the reflection features observed in the X-ray 
 range, i.e. 
the fluorescence FeK line and the hump. In particular it can explain
the broad profile of the Fe line (with extended blue and red
 wings), which in this case is due to comptonization, without requiring
a relativistic broadening by
reflection on a thin disc located very close to the BH (Abrassart \& Dumont 
1998, 
 1999). 

\section{Third possibility: completely non standard discs}

If the release of gravitational energy is not local, the emitted spectrum 
can be completely different from that of a standard disc. 
The energy can be transported magnetically and released far 
from the BH, for instance in the form of relativistic particles emitting 
synchrotron radiation or through shocks formed in jets. In this case the
emission spectrum depends on the mechanism of energy release, and is not 
known. 
Another possibility is that in the remote regions of the disc (which are 
self-gravitating, see Collin \& Hur\'e 2001 for the Kaspi et al. 
sample), self-gravitation induces changes in the rotation curve, which 
could become 
non-Keplerian. Such a mechanism has been proposed
in the context of protostellar discs to explain their infrared spectrum 
(Lodato \& Bertin 2001). Since these models are presently not sufficiently 
elaborated in the framework of AGN, we leave this question opened.

\section{Conclusion}

	Though known with large uncertainties, the BH masses deduced from 
reverberation studies are now proved to be
consistent with the relationship between BH mass and galaxy velocity 
dispersion (Laor 1998, 2001). For the sample of 34 AGN studied by reverberation mapping,
 we have used these masses to deduce 
the accretion rates and the corresponding 
bolometric luminosities, assuming that the observed optical 
luminosity is provided by the accretion disc. Through a discussion of the different sources of 
uncertainties, we have shown that it is difficult to escape the 
conclusion that, {\it either} a large fraction of the objects are accreting at 
Eddington ou super-Eddington rates through geometrically thick disc, {\it or} the
 optical luminosity is not 
provided by the disc. 

We summarize here the main conclusions.

\noindent 1. If a fraction of the objects are accreting at super-Eddington rates,
their bolometric luminosity is limited to the Eddington luminosity, and 
the ratio of their optical to bolometric luminosity agrees with the observed 
ratio $\sim 10$.  As already known, we show that in general it is difficult to account
 for the
 observed spectrum in the optical-UV range with a standard accretion disc, 
 in particular for the super-Eddington objects: 
the observed spectrum is too ``soft", or in other words, there is too much 
energy observed
in the optical band compared to the UV
band. It is necessary to assume that the disc is truncated at an inner radius 
much 
larger than the last stable orbit, or for any reason does not emit UV radiation. On the 
other hand, present X-ray 
observations are finally not able to help dismiss the ``super-Eddington 
solution".

\noindent 2. If a large fraction of AGN are accreting at super-Eddington rates (via 
geometrically thick discs), some important evolutionary and 
cosmological consequences follow. 

{\it First} the most massive 
objects are accreting at the smallest rates (expressed in Eddington ratios), 
i.e.  the accretion rate is about inversely proportional 
to the mass, while an opposite result is found if the accretion 
rate is assumed to be proportional to the optical luminosity, as always done.

{\it Second}
the lifetime of these AGN should be very short, and since the efficiency of 
mass-energy conversion is low in this
process,  the 
mass density locked in dead quasars should be larger than it is deduced 
for a ``normal" efficiency. 

Incidentally we find also that the NLS1 nuclei are at 
the extreme of a well defined sequence relating the Eddington ratio to the 
line widths. 

\noindent 3. If the optical luminosity is not emitted by a standard disc, 
super-Eddington rates are not required. However, in order to get 
a featureless emission like the blue bump, it is 
necessary to appeal to a dense and optically thick medium, and the 
observed luminosities imply a very large emitting surface, therefore large 
distances from the center. Either a warped geometrically thin disc 
illuminated by the central 
X-ray source, or a 
quasi-spherical system of dense 
clouds,  can constitute an adequate emitting medium.
 In the latter case, such a system of clouds 
 might also be 
responsible for the reflection features in 
the X-ray range, i.e. the broad Fe K line and the hump.
Though these dense 
 clouds can participate to the 
 accretion flow, a tempting hypothesis is that they pertain to
an outflowing wind,  whose 
dilute counterpart constitutes the Warm Absorber and the Broad Absorption Line
 region. 

\bigskip

In conclusion  we are not able to give a firm answer to the question raised 
 in the title. It is quite possible that a large fraction of objects 
 accrete at super-Eddington rates, with all the consequences that it would 
 imply, or that the Blue Bump - at least its optical 
 part - is not due to a standard disc, and thus does not require super-Eddington 
 accretion rates.

{}

\end{document}